\documentclass[aps,showpacs,pra,superscriptaddress,]{revtex4}

\usepackage{amsmath}
\usepackage{amsfonts}
\usepackage{amssymb}
\usepackage{bm}
\usepackage{graphicx}

\begin{document}

\title{Solitons and periodic wave solutions for complex Ginzburg-Landau
equation modelling fiber lasers and nonequilibrium phenomena}
\author{Vladimir I. Kruglov}
\affiliation{Centre for Engineering Quantum Systems, School of Mathematics and Physics,
The University of Queensland, Brisbane, Queensland 4072, Australia}
\author{ Houria Triki}
\affiliation{Radiation Physics Laboratory, Department of Physics, Faculty of Sciences,
Badji Mokhtar University, P. O. Box 12, 23000 Annaba, Algeria}

\begin{abstract}
New types of soliton and periodic waves are identified for a nonlinear
dissipative medium where the pulse propagation is governed by the cubic
complex Ginzburg-Landau equation. We find that the dynamical equation for
the pulse amplitude supports two distinct types of kink and antikink
solitons with different functional forms. It is found that the obtained kink
and antikink soliton waveforms occur under the same fixed inverse velocity.
The results also indicate that the periodic waves can propagate with variety
of wave forms such as \textrm{sn}, \textrm{cn}, \textrm{dn} and their
rational forms as well. It is also shown that in the long-wave limit, the
derived periodic waves degenerate into different bright and dark soliton
pulses. The stability analysis based on the theory of dispersive waves in
nonlinear optics is developed. It is shown that some elliptic and soliton
solutions are quasi-stable in the context of passive mode locking lasers
described by complex Ginzburg-Landau equation.
\end{abstract}

\maketitle

\section{Introduction}

Transmission of optical solitons in nonlinear media has attracted
considerable attention motivated by their important potential applications
to high-capacity fiber optical communications and to all-optical switching 
\cite{Agraw,Haseg}. A fundamental property of solitons is their ability to
propagate over long distances without spreading. The formation of such
shape-preserving wave packets in optical fibers results from an exact
balance between group velocity dispersion and self-phase modulation. For the
anomalous dispersion regime, the bright soliton pulse is formed in the fiber 
\cite{Hasegawa1}, whereas the dark soliton is generated in the normal
dispersion regime \cite{Hasegawa2}. Experimentally, the observation of
envelope solitons has been demonstrated in different nonlinear media,
including monomode fibers \cite{Mollenauer}, bulk optical materials \cite%
{Aitchison}, femtosecond lasers \cite{Salin}, and photorefractive crystals 
\cite{Photo1,Photo2}.

Continuous interest has focused on the so-called dissipative solitons, which
arise in fiber lasers due to the mutual interaction among the cavity
dispersion, laser gain saturation, fiber nonlinearity, and gain bandwidth 
\cite{Song}. The experimental observation of such kind of solitons has been
demonstrated in \cite{R7}. A key model used to describe the propagation of
these solitons is the complex Ginzburg-Landau equation \cite{R1,R2}, which
also models a wide variety of nonlinear phenomena across various fields of
physics, ranging from superconductivity \cite{R3}, superfluidity \cite{R4},
to Bose-Einstein condensation \cite{R5} and liquid crystals \cite{R6}.

Studying closed form solutions of the complex Ginzburg-Landau equation is of
great physical relevance due to their essential role in describing a wide
variety of nonlinear phenomena, such as dissipative solitons, pulsating
waves, and pattern formation in non-equilibrium systems. It is interesting
to note that in addition to stationary solitons, sinks, sources, moving
solitons and fronts \cite{R81,R82,R83}, the variants of the complex
Ginzburg-Landau equation have been found to exhibit important localized
solutions, such as pulsating, erupting, creeping, and chaotic solitons \cite%
{R91,R92}.

It is critical to examine whether novel localized and periodic structures
can exist in nonlinear dissipative systems governed by the cubic complex
Ginzburg-Landau equation (CGLE). Identifying different types of nonlinear
waves of this pertinent model is of profound theoretical and practical
significance. In this work, we demonstrate that envelope solitons and
periodic waves of various kinds are possible in such nonlinear media. The
characteristics of these nonlinear waves are discussed and their dynamics is
also analyzed.

The paper consists of the following. In Sec. II, we show that the CGLE
possesses two distinct types of analytical kink and antikink soliton
solutions which take different functional forms and propagate with fixed
inverse velocity. In Sec. III, we introduce novel periodical wave solutions
and find the corresponding bright and dark solitons of the model equation at
the long-wave limit. In Sec. IV, we examine the stability analysis of the
newly identified periodic and soliton solutions based on the theory of
optical nonlinear dispersive waves. The conclusions of this paper will then
be summarized in Section V.

\section{Ginzburg-Landau equation and kink antikink wave solutions}

In the context of fiber lasers, the Ginzburg-Landau equation \cite{Agraw} is
given as 
\begin{equation}
i\frac{\partial \psi }{\partial z}+(\alpha -i\beta )\frac{\partial ^{2}\psi 
}{\partial t^{2}}+(\gamma -i\sigma )\left\vert \psi \right\vert ^{2}\psi
-i\delta \psi =0,  \label{1}
\end{equation}%
where $\psi $ is the normalized envelope of the electric field, $z$ and $t$
are the propagation distance and retarded time, $\alpha $ accounts for the
group delay dispersion coefficient which is positive and negative for
anomalous and normal dispersion regimes respectively, $\beta $ describes the
spectral filtering, $\gamma $ is the Kerr-nonlinear coefficient, $\sigma $
is the nonlinear gain or absorption coefficient, $\delta $ represents the
linear gain or loss. The coefficients in this equation are connected with
the parameters of model for passive mode locking fiber lasers based on the
Ginzburg-Landau equation. We present the definition of these parameters for
passive mode locking fiber lasers in Appendix A.

A challenging problem in nonlinear dynamics is the identification of new
classes of soliton waveforms that may appear when considering the
contribution of all physical mechanisms in the system. This is because
soliton solutions when they exist in analytic form are very attractive to
compare experimental results with theory \cite{Chab}. Moreover, obtaining
exact soliton solutions enables us to determine relevant physical quantities
analytically and serve as diagnostics for simulations \cite{Cooper}. In the
following, new types of kink and antikink soliton solutions are identified
in a fiber laser, wherein the light propagation is modeled by the CGLE (\ref%
{1}).

To obtain the exact kink, antikink, periodic and soliton solutions of Eq. (%
\ref{1}), we consider the following complex envelope for traveling-wave
solutions: 
\begin{equation}
\psi (z,t)=F(\xi )\exp \left[ i(\kappa z-\omega_{s} t+\theta )\right] ,
\label{2}
\end{equation}%
where $\xi =t-uz$ is the traveling coordinate and $F$ is a real function of $%
\xi $. Here $u=1/\mathrm{v}$, with \textrm{v} the group velocity of the wave
packet, $\kappa $ is the wave number, $\omega_{s} $\ represents the
frequency shift, and $\theta $ is the phase constant.

Substituting Eq. (\ref{2}) into (\ref{1}) and separating real and imaginary
parts, the following coupled equations result:%
\begin{equation}
F^{\prime \prime }-\frac{2\beta }{\alpha }\omega_{s} F^{\prime }-\left( 
\frac{\kappa }{\alpha }+\omega_{s} ^{2}\right) F+\frac{\gamma }{\alpha }%
F^{3}=0,  \label{3}
\end{equation}%
\begin{equation}
F^{\prime \prime }+\left( \frac{u}{\beta }+\frac{2\alpha \omega_{s} }{\beta }%
\right) F^{\prime }+\left( \frac{\delta }{\beta }-\omega_{s} ^{2}\right) F+%
\frac{\sigma }{\beta }F^{3}=0.  \label{4}
\end{equation}%
Equations (\ref{3}) and (\ref{4}) will be equivalent, provided that the
following conditions are satisfied: 
\begin{equation}
-\frac{2\beta }{\alpha }\omega_{s} =\frac{u}{\beta }+\frac{2\alpha
\omega_{s} }{\beta },~~~~\frac{\kappa }{\alpha }+\frac{\delta }{\beta }%
=0,~~~~\frac{\gamma }{\alpha }=\frac{\sigma }{\beta }.  \label{5}
\end{equation}%
Hence, we have the parameters $\kappa $ and $\omega_{s} $ as 
\begin{equation}
\kappa =-\frac{\alpha \delta }{\beta },~~~~\omega_{s} =-\frac{\alpha u}{%
2(\alpha ^{2}+\beta ^{2})},  \label{6}
\end{equation}%
where $u=1/v$ is a free parameter. Moreover, the third equation in Eq. (\ref%
{5}) yields the constraint condition as $\beta \gamma =\alpha \sigma $.
Thus, the system of Eqs. (\ref{3}) and (\ref{4}) reduces to one equation as 
\begin{equation}
F^{\prime \prime }+aF^{\prime }+bF+cF^{3}=0,  \label{7}
\end{equation}%
where the coefficients $a$, $b$ and $c$ are given as 
\begin{equation}
a=\frac{\beta u}{\alpha ^{2}+\beta ^{2}},\quad b=\frac{\delta }{\beta }-%
\frac{\alpha ^{2}u^{2}}{4(\alpha ^{2}+\beta ^{2})^{2}},\quad c=\frac{\sigma}{%
\beta }.  \label{8}
\end{equation}

We have found two types of kink and antikink solutions of Eq. (\ref{7}) with
condition as $a^{2}=9b/2$ which is equivalent to fixed inverse velocity $u$
in Eq. (\ref{8}) as 
\begin{equation}
u=\pm 6(\alpha ^{2}+\beta ^{2})\sqrt{\frac{\delta }{\beta
(9\alpha^{2}+8\beta ^{2})}},  \label{9}
\end{equation}%
The parameters in Eq. (\ref{8}) and the frequency $\omega_{s}$ for the
inverse velocity given by Eq. $(\ref{9})$ are 
\begin{equation}
a=\pm 6\beta\sqrt{\frac{\delta }{\beta (9\alpha^{2} +8\beta ^{2})}},~~~~b=%
\frac{8\beta\delta}{9\alpha^{2} +8\beta ^{2}},~~~~c=\frac{\sigma}{\beta }%
,~~~~ \omega_{s} =\mp 3\alpha \sqrt{\frac{\delta }{\beta (9\alpha^{2}+8\beta
^{2})}}.  \label{10}
\end{equation}%
We present below two types of kink and antikink solutions for Eq. (\ref{7})
with fixed inverse velocity $u$ given by Eq. (\ref{9}), and the kink and
antikink solution with the inverse velocity $u=0$.

\begin{figure}[h]
\includegraphics[width=1.3\textwidth]{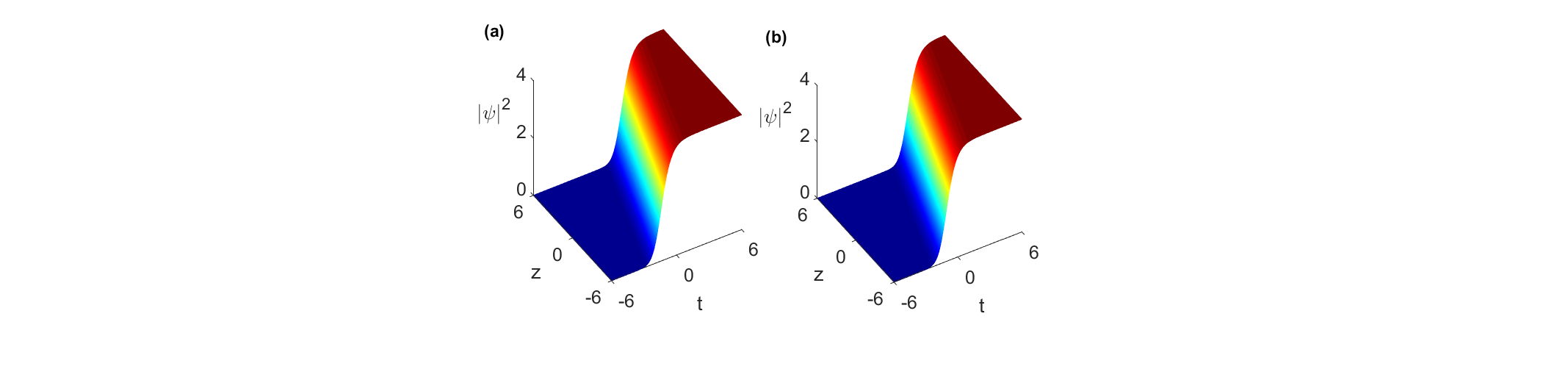}
\caption{Evolution of soliton solutions with parameters $\protect\alpha %
=0.32 $, $\protect\beta =0.35$, $\protect\gamma =-0.64$, $\protect\sigma %
=-0.7$, $\protect\delta =5.434$, $\protect\xi_{0}=0$ (a) soliton solution (%
\protect\ref{13}) with $\protect\nu =1$ (b) soliton solution (\protect\ref%
{16}) with the $-$ sign of $\protect\lambda $ when $S=2.$}
\label{FIG.1.}
\end{figure}

\begin{figure}[h]
\includegraphics[width=1.3\textwidth]{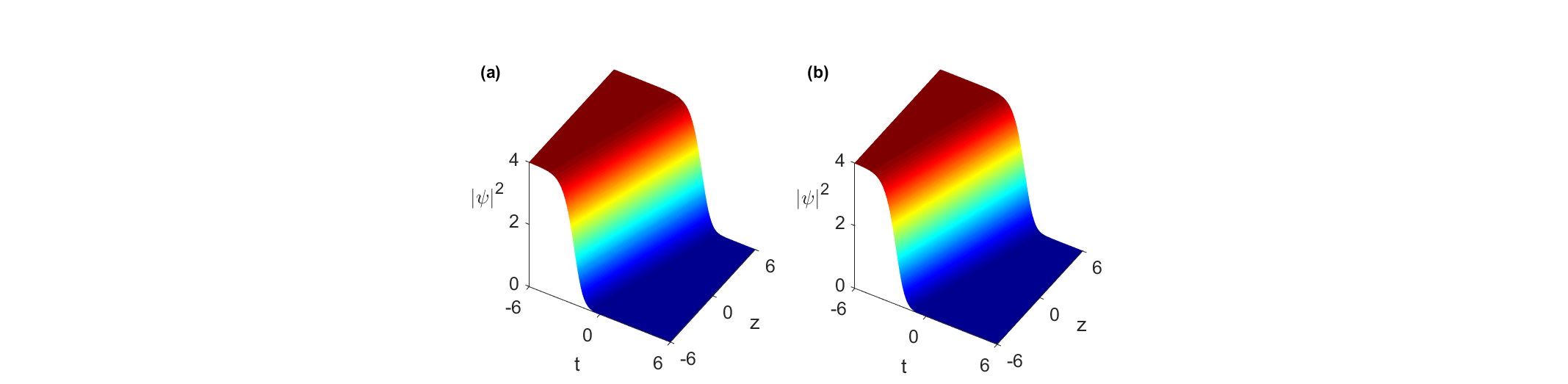}
\caption{Evolution of soliton solutions with parameters $\protect\alpha %
=0.32 $, $\protect\beta =0.35$, $\protect\gamma =-0.64$, $\protect\sigma %
=-0.7$, $\protect\delta =5.434$, $\protect\xi_{0}=0$ (a) soliton solution (%
\protect\ref{13}) with $\protect\nu =-1$ (b) soliton solution (\protect\ref%
{16}) with the $+$ sign of $\protect\lambda $ when $S=2$.}
\label{FIG.2.}
\end{figure}

\begin{figure}[h]
\includegraphics[width=1\textwidth]{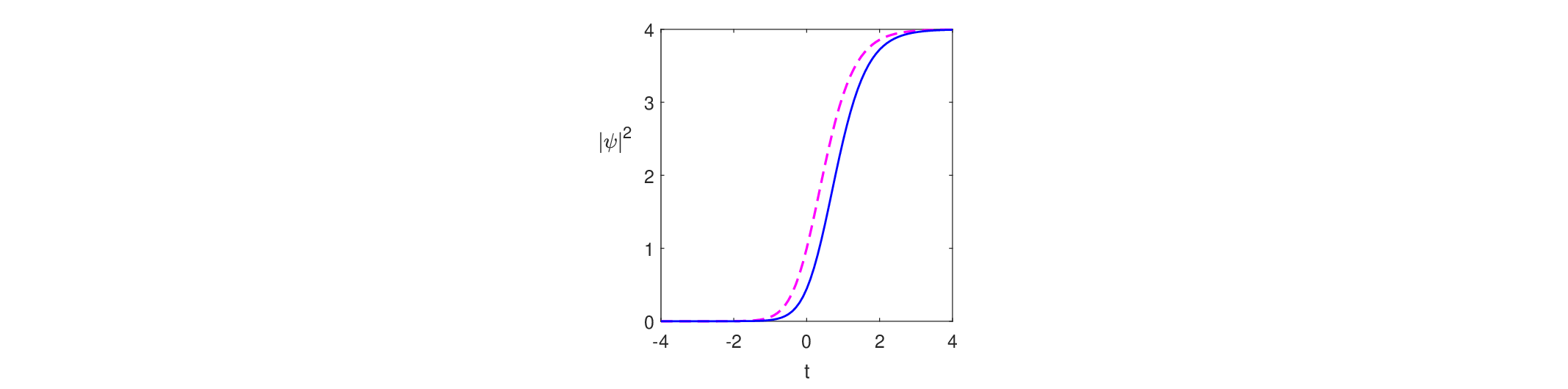}
\caption{Intensity profiles of kink soliton solution (\protect\ref{13}) with 
$\protect\nu=1$ (dashed line) and kink soliton solution (\protect\ref{16})
with the $-$ sign of $\protect\lambda $ (thick line). The other parameters
are same as in Fig. 1.}
\label{FIG.3.}
\end{figure}

\begin{figure}[h]
\includegraphics[width=1\textwidth]{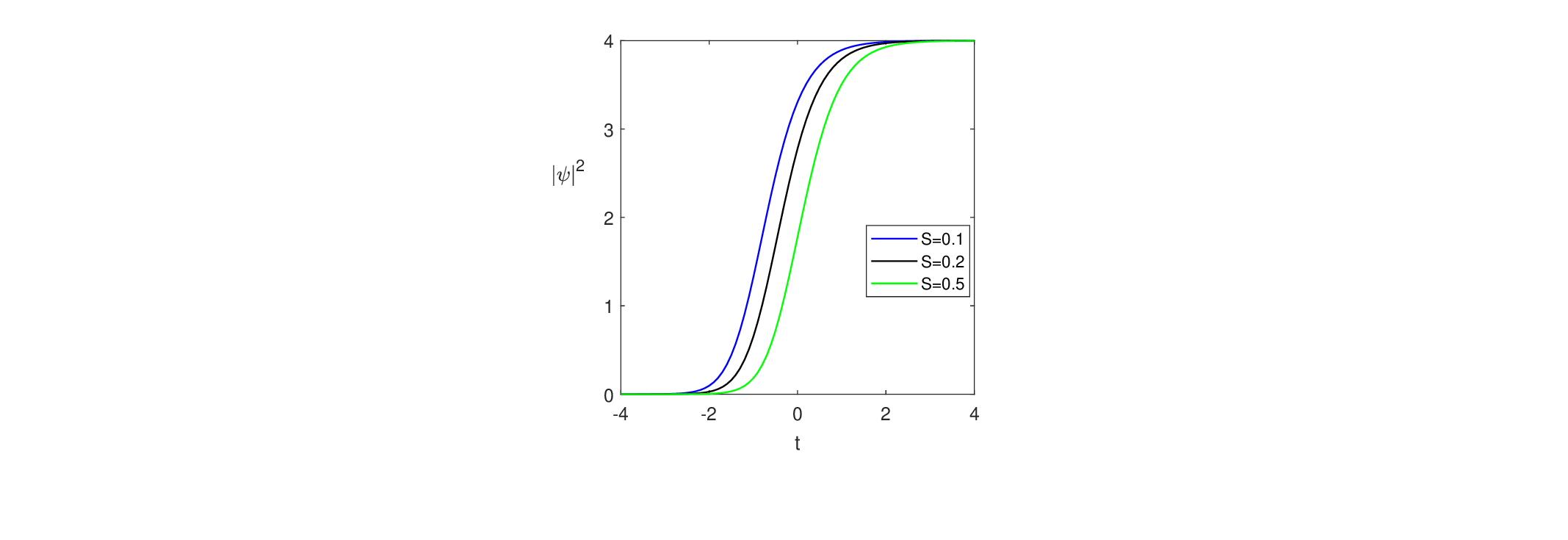}
\caption{Intensity profiles of kink soliton solution (\protect\ref{16}) with
distinct values of $S$. The other parameters are same as in Fig. 1(c).}
\label{FIG.4.}
\end{figure}

\begin{description}
\item[\textbf{1. First type solution}] 
\end{description}

We have found a first type of exact analytical solution for Eq. (\ref{7})
with the fixed inverse velocity $u$ given by Eq. (\ref{9}): 
\begin{equation}
F\left( \xi \right) =A\left\{ 1+\nu \,\mathrm{tanh}[\mu (\xi -\xi
_{0})]\right\} ,  \label{11}
\end{equation}%
with $\nu ^{2}=1$, $A=\pm \sqrt{-b/4c}$ and $\mu =\sqrt{b/8}$. Thus, using
Eq. (\ref{10}) we have the amplitude and inverse width as 
\begin{equation}
A=\pm \sqrt{-\frac{2\delta \beta ^{2}}{\sigma (9\alpha ^{2}+8\beta ^{2})}}%
,\quad \mu =\sqrt{\frac{\delta \beta }{9\alpha ^{2}+8\beta ^{2}}}.
\label{12}
\end{equation}%
Hence, we have obtained an exact solution for the CGLE (\ref{1}) as%
\begin{equation}
\psi (z,t)=A\left\{ 1+\nu \,\mathrm{tanh}\left[ \mu \left( \xi -\xi
_{0}\right) \right] \right\} \exp \left[ i(\kappa z-\omega _{s}t+\theta )%
\right] ,  \label{13}
\end{equation}%
where $\nu =1$ and $\nu =-1$ for different solutions. This solution takes
place for material parameters obeying the conditions $\delta \beta >0$ and $%
\sigma \delta <0$.

\begin{description}
\item[\textbf{2. Second type solution}] 
\end{description}

We also found that Eq. (\ref{7}) admits a second type of exact analytical
solution with the fixed inverse velocity $u$ given by Eq. (\ref{9}): 
\begin{equation}
F\left( \xi \right) =\frac{\Lambda }{1+S\,\mathrm{\exp }[\lambda (\xi -\xi
_{0})]},  \label{14}
\end{equation}%
with the amplitude $\Lambda =\pm \sqrt{-b/c}$ and inverse width $\lambda
=a/3 $. The second type bounded solution exists for an arbitrary positive ($%
S\,>0$) parameter $S\,$. Using Eq. (\ref{10}) we have the amplitude and
inverse width as 
\begin{equation}
\Lambda =\pm \sqrt{-\frac{8\delta \beta ^{2}}{\sigma (9\alpha ^{2}+8\beta
^{2})}},\quad \lambda =\pm 2\beta \sqrt{\frac{\delta }{\beta (9\alpha
^{2}+8\beta ^{2})}}.  \label{15}
\end{equation}%
Hence, Eq. (\ref{2}) leads to solution for the CGLE (\ref{1}) as 
\begin{equation}
\psi (z,t)=\frac{\Lambda }{1+S\,\mathrm{\exp }\left[ \lambda \left( \xi -\xi
_{0}\right) \right] }\exp \left[ i(\kappa z-\omega _{s}t+\theta )\right] ,
\label{16}
\end{equation}%
\noindent provided that $\delta \beta >0$ and $\sigma \delta <0$. The
bounded solution of this type exists for an arbitrary positive parameter $%
S\, $.

\begin{description}
\item[\textbf{3. Third type solution}] 
\end{description}

We have also found that Eq. (\ref{7}) admits a third type of exact solution
in the case when the inverse velocity $u=0$. In this case we have $a=0$, $%
b=\delta/\beta$, $c=\sigma/\beta$ and the kink wave solution is 
\begin{equation}
\psi (z,t)=B\,\mathrm{tanh}(w(t -t _{0}))\exp \left[ i(\kappa z+\theta )%
\right] ,  \label{17}
\end{equation}%
with the amplitude $B$ and inverse width $w$ as 
\begin{equation}
B=\pm \sqrt{-\frac{\delta}{\sigma}},\quad w =\sqrt{\frac{\delta}{2\beta}},
\label{18}
\end{equation}
provided that $\delta\beta>0$ and $\delta\sigma<0$.

Figure 1(a) shows the evolution of the intensity profile of the soliton
solution (\ref{13}) with $\nu =1$ for the parameter values: $\alpha =0.32$, $%
\beta =0.35$, $\gamma =-0.64$, $\sigma =-0.7$, $\delta =5.434$, and $\xi
_{0}=0.$ The result for the soliton solution (\ref{16}) taking the negative
value of $\lambda $ is illustrated in Fig. 1(b). From these figures, we
observe that the two types of soliton solutions exhibit a kink-type wave
profile. Furthermore, we notice that the soliton solution (\ref{13}) with $%
\nu =-1$ as well as the pulse represented by the solution (\ref{16}) for the
positive value of $\lambda $ are antikink type waves, as depicted in Fig. 2.

Figure 3 shows the comparison of two types of kink solitons with choice of
the same parameter values as those in Fig. 1$.$ It can be clearly seen that
the kink soliton waveform (\ref{16}) is more steep than the one given in (%
\ref{13}).

In Fig. 4, we illustrate the intensity profile of kink soliton waveform (\ref%
{16}) for three different values of parameter $S$, $0.1$, $0.2$ and $0.5$.
We can see from this figure that the soliton pulse shifts slightly as the
parameter $S$ increases but the shape of the soliton remain unchanged. This
indicates that the shape of kink wave can be controlled by varying this
parameter.

\section{Periodic, kink and soliton solutions of Ginzburg-Landau equation}

We consider in this section the solutions of Eq. (\ref{7}) for two cases:
when inverse velocity is $u=0$ or the inverse velocity is mach smaller of
some characteristic parameter. We can find appropriate condition for free
parameter $u$ rewriting Eq. (\ref{7}) in dimensionless form. Let us define
the dimensionless variable $s=w\xi $ and dimensionless function $f(s)=F(\xi
)/F_{0}$ where $w$ and $F_{0}$ are the characteristic inverse time and the
amplitude of pulses respectively. The parameters $w$ and $F_{0}=B$ are found
below for different solutions of Eq. (\ref{7}). The dimensionless form of
Eq. (\ref{7}) is given as 
\begin{equation}
\frac{d^{2}f}{ds^{2}}+\frac{a}{w}\frac{df}{ds}+\frac{b}{w^{2}}f+\frac{%
cF_{0}^{2}}{w^{2}}f^{3}=0.  \label{19}
\end{equation}%
We consider the condition $|a|/w\ll 1$ which can also be written as 
\begin{equation}
|\beta u|\ll w(\alpha ^{2}+\beta ^{2}).  \label{20}
\end{equation}%
Thus, when this condition for free parameter $u$ is satisfied one can
neglect the second term (proportional to the derivative $df/ds$) in Eq. (\ref%
{19}). In this case Eq. (\ref{7}) reduces to the following nonlinear
differential equation: 
\begin{equation}
F^{\prime \prime }+bF+cF^{3}=0.  \label{21}
\end{equation}%
Integration of this equation yields the first order nonlinear differential
equation: 
\begin{equation}
\left( \frac{dF}{d\xi }\right) ^{2}+bF^{2}+\frac{c}{2}F^{4}=Q,  \label{22}
\end{equation}%
where $Q$ is an integration constant.

We consider below the solutions of Eq. (\ref{1}) based on Eq. (\ref{22})
with the condition for inverse velocity $u$ given in Eq. (\ref{20}). In the
limiting case when $u=0$ we have the parameter $a=0$ and hence Eq. (\ref{7})
yields Eq. (\ref{22}). In this limiting case we have $b=\delta /\beta $, $%
c=\sigma /\beta $, $\omega =0$ and the variable $\xi =t$.

\begin{description}
\item[\textbf{1. Periodic sn(x,k) and tanh(x)} solutions] 
\end{description}

Equation (\ref{22}) has the elliptic solution: 
\begin{equation}
F(\xi )=B\,\mathrm{sn}(w(\xi -\xi _{0}),k).  \label{23}
\end{equation}
Equations (\ref{22}) and (\ref{23}) yield the following equation: 
\begin{equation}
B^{2}w^{2}-B^{2}w^{2}(1+k^{2})\mathrm{sn}^{2}(x)+B^{2}w^{2}k^{2}\mathrm{sn}%
^{4}(x)+bB^{2}\mathrm{sn}^{2}(x)+\frac{c}{2}B^{4}\mathrm{sn}^{4}(x)=Q,
\label{24}
\end{equation}%
where $x=w(\xi -\xi _{0})$ and $\mathrm{sn}(x)\equiv \mathrm{sn}(x,k)$ is
the Jacobi elliptic function with the modulus $k$. This equation yields the
integration constant as $Q=B^{2}w^{2}$, then Eq. (\ref{24}) leads to inverse
width $w$ and amplitude $B$ as 
\begin{equation}
w=\sqrt{\frac{b}{1+k^{2}}},~~~~B=\pm k\sqrt{-\frac{2b}{c(1+k^{2})}},
\label{25}
\end{equation}%
where $b>0$ and $c<0$. The elliptic solution given in Eq. (\ref{23}) and Eq.
(\ref{2}) leads to periodic wave function for the CGLE (\ref{1}) as 
\begin{equation}
\psi (z,t)=\pm k\sqrt{-\frac{2b}{c(1+k^{2})}}\,\mathrm{sn}(w(\xi -\xi
_{0}),k)\exp \left[ i(\kappa z-\omega_{s} t+\theta )\right] ,  \label{26}
\end{equation}%
where the inverse width $w$ is given in Eq. (\ref{25}). Let us choose $k=1$
then Eq. (\ref{26}) yields the kink wave solution: 
\begin{equation}
\psi (z,t)=\pm \sqrt{-\frac{b}{c}}\,\mathrm{tanh}(w(\xi -\xi _{0}))\exp %
\left[ i(\kappa z-\omega_{s} t+\theta )\right] ,  \label{27}
\end{equation}%
with $w=\sqrt{b/2}$, and $b>0$, $c<0$. 
\begin{figure}[h]
\includegraphics[width=1.3\textwidth]{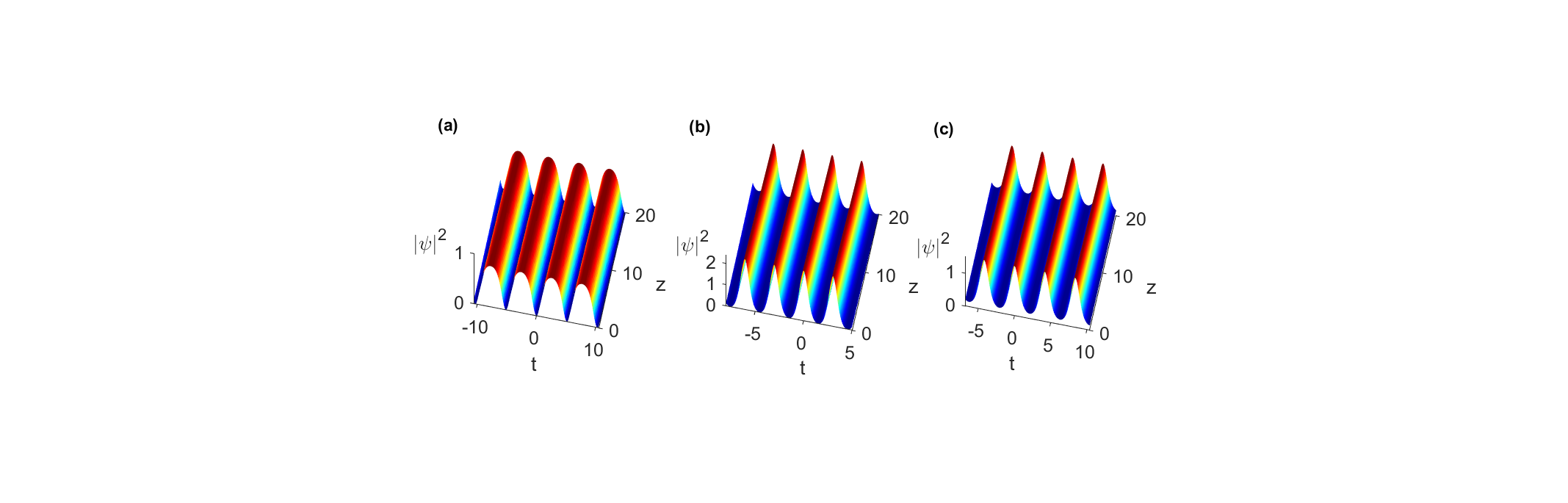}
\caption{Evolution of periodic wave solutions with parameters $\protect%
\alpha =0.2$, $\protect\beta =0.1$, $\protect\xi _{0}=0$, $k=0.9$, $u=0.01$
(a) periodic wave solution (\protect\ref{26}) with $\protect\gamma =-0.4$, $%
\protect\sigma =-0.2$, $\protect\delta =0.182$ (b) periodic wave solution (%
\protect\ref{31}) with $\protect\gamma =0.4$, $\protect\sigma =0.2$, $%
\protect\delta=-0.182$ (c) periodic wave solution (\protect\ref{37}) with $%
\protect\gamma =0.4$, $\protect\sigma =0.2$, $\protect\delta =-0.182$.}
\label{FIG.5.}
\end{figure}

\begin{figure}[h]
\includegraphics[width=1.3\textwidth]{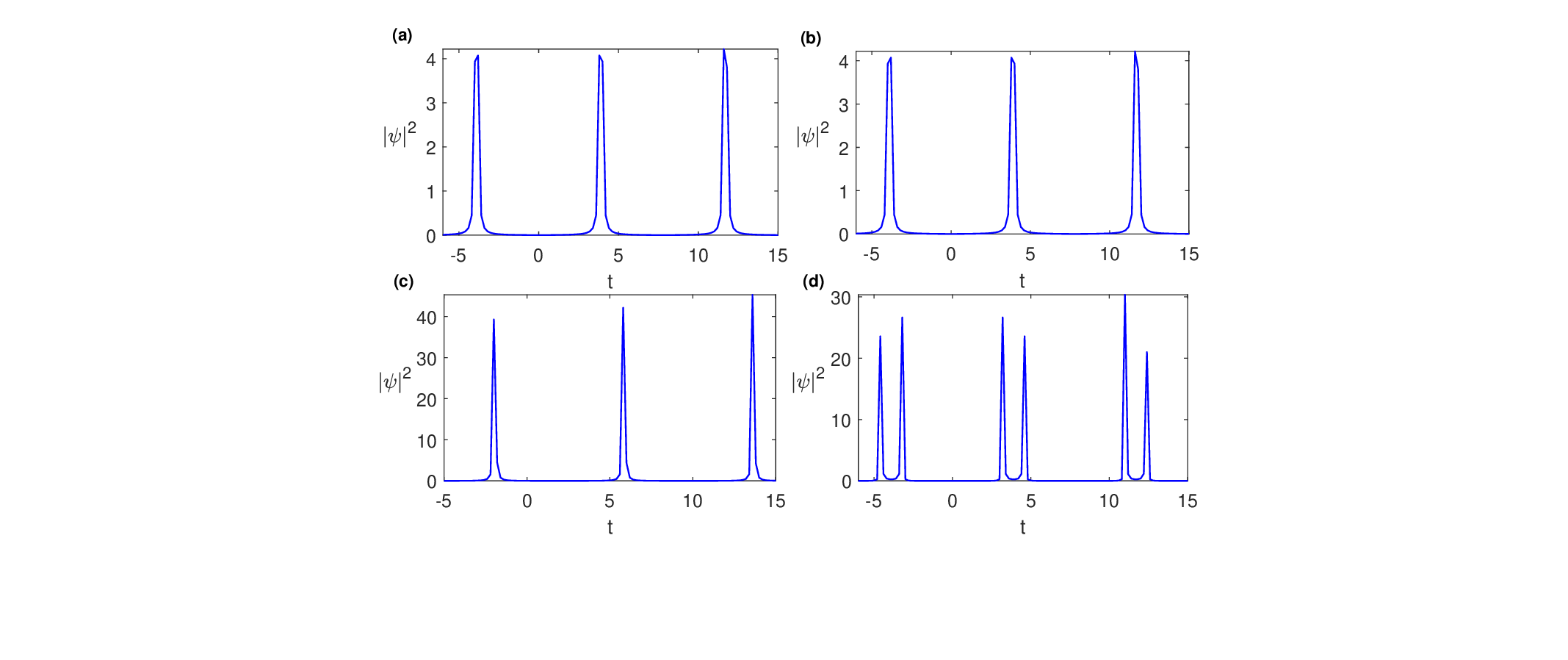}
\caption{Evolution of periodic wave solutions with parameters $\protect%
\alpha =0.2$, $\protect\beta =0.1$, $\protect\xi _{0}=0$, $k=0.6$, $u=0.01$
(a) periodic wave solution (\protect\ref{42}) with $\protect\gamma =-10$, $%
\protect\sigma =-5$, $\protect\delta =0.014$ (b) periodic wave solution (%
\protect\ref{48}) with $\protect\gamma =-0.326$, $\protect\sigma =-0.163$, $%
\protect\delta=0.083$ (c) periodic wave solution (\protect\ref{53}) with $%
\protect\gamma =6.4$, $\protect\sigma =3.2$, $\protect\delta =-0.068$ (d)
periodic wave solution (\protect\ref{58}) with $\protect\gamma =-0.326$, $%
\protect\sigma =-0.163$, $\protect\delta =0.083$.}
\label{FIG.6.}
\end{figure}

\begin{figure}[h]
\includegraphics[width=1.3\textwidth]{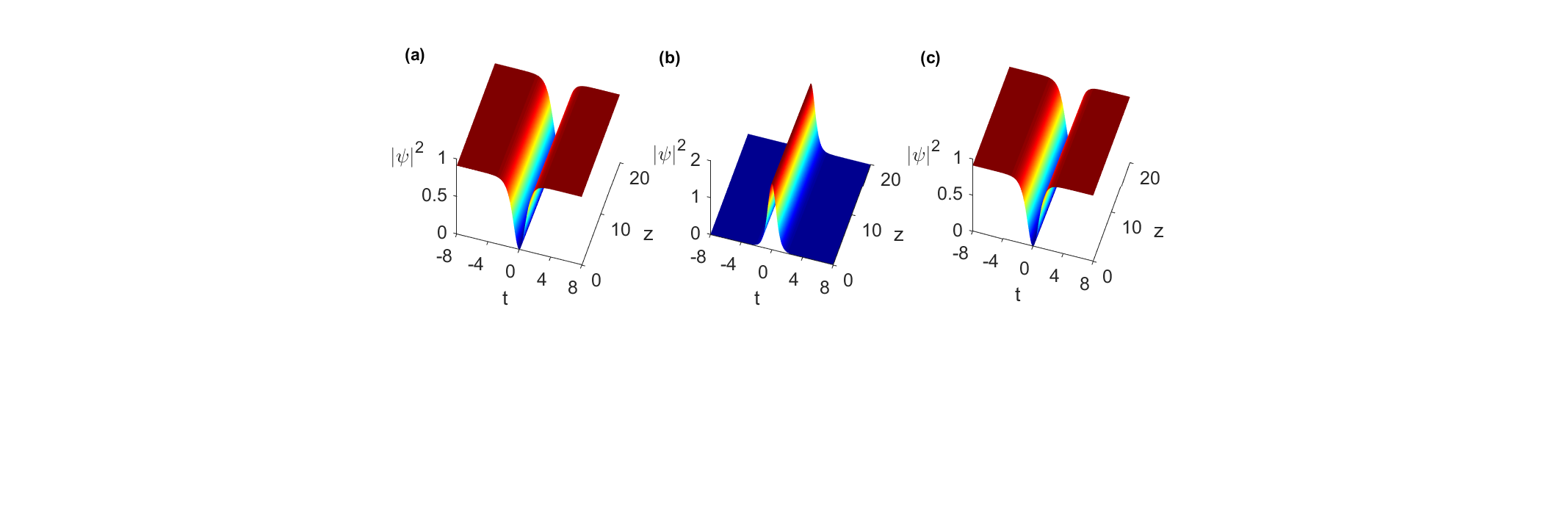}
\caption{Evolution of soliton solutions (a) dark soliton solution (\protect
\ref{27}) (b) bright soliton solution (\protect\ref{32}) and (c) dark
soliton solution (\protect\ref{43}) for the values mentioned in the text.}
\label{FIG.7.}
\end{figure}

\begin{description}
\item[\textbf{2. Periodic cn(x,k) and sech(x) solutions}] 
\end{description}

Equation (\ref{22}) has the elliptic solution: 
\begin{equation}
F(\xi )=B\,\mathrm{cn}(w(\xi -\xi _{0}),k).  \label{28}
\end{equation}%
Equations (\ref{22}) and (\ref{28}) yield the following equation: 
\begin{equation}
B^{2}w^{2}(1-k^{2})+B^{2}w^{2}(2k^{2}-1)\mathrm{cn}^{2}(x)-B^{2}w^{2}k^{2}%
\mathrm{cn}^{4}(x)+bB^{2}\mathrm{cn}^{2}(x)+\frac{c}{2}B^{4}\mathrm{cn}%
^{4}(x)=Q,  \label{29}
\end{equation}%
where $\mathrm{cn}(x)\equiv \mathrm{cn}(x,k)$ is the Jacobi elliptic
function with the modulus $k$. This equation leads to integration constant
as $Q=B^{2}w^{2}(1-k^{2})$, hence the inverse width $w$ and amplitude $B$
are 
\begin{equation}
w=\sqrt{\frac{b}{1-2k^{2}}},~~~~B=\pm k\sqrt{\frac{2b}{c(1-2k^{2})}},
\label{30}
\end{equation}%
where $b>0$ for $1-2k^{2}>0$ and $b<0$ for $1-2k^{2}<0$ and $c>0$ for all $k$%
. The elliptic solution given in Eq. (\ref{28}) and Eq. (\ref{2}) lead to
periodic wave function for the Ginzburg-Landau Eq. (\ref{1}) as 
\begin{equation}
\psi (z,t)=\pm k\sqrt{\frac{2b}{c(1-2k^{2})}}\,\mathrm{cn}(w(\xi -\xi
_{0}),k)\exp \left[ i(\kappa z-\omega_{s} t+\theta )\right] ,  \label{31}
\end{equation}%
where the inverse width $w$ is given in Eq. (\ref{30}). Let us choose $k=1$
then Eq. (\ref{31}) yields the soliton wave solution: 
\begin{equation}
\psi (z,t)=\pm \sqrt{-\frac{2b}{c}}\,\mathrm{sech}(w(\xi -\xi _{0}))\exp %
\left[ i(\kappa z-\omega_{s} t+\theta )\right] ,  \label{32}
\end{equation}%
with $w=\sqrt{-b}$, and $b<0$, $c>0$. This soliton solution has the
following energy integral, 
\begin{equation}
\mathcal{E}=\int_{-\infty }^{+\infty }|\psi (z,t)|^{2}dt=4\sqrt{-\frac{b}{%
c^{2}}}.  \label{33}
\end{equation}

\begin{description}
\item[\textbf{3. Periodic dn(x,k) and sech(x) solutions}] 
\end{description}

Equation (\ref{22}) has the elliptic solution: 
\begin{equation}
F(\xi )=B\,\mathrm{dn}(w(\xi -\xi _{0}),k).  \label{34}
\end{equation}%
Equation (\ref{22}) with Eq. (\ref{34}) yields the following equation: 
\begin{equation}
B^{2}w^{2}(k^{2}-1)+B^{2}w^{2}(2-k^{2})\mathrm{dn}^{2}(x)-B^{2}w^{2}\mathrm{%
dn}^{4}(x)+bB^{2}\mathrm{dn}^{2}(x)+\frac{c}{2}B^{4}\mathrm{dn}^{4}(x)=Q,
\label{35}
\end{equation}%
were $\mathrm{dn}(x)\equiv \mathrm{dn}(x,k)$ is the Jacobi elliptic function
with the modulus $k$. This equation yields the integration constant as $%
Q=B^{2}w^{2}(k^{2}-1)$, hence the inverse width $w$ and amplitude $B$ are 
\begin{equation}
w=\sqrt{\frac{b}{k^{2}-2}},~~~~B=\pm \sqrt{\frac{2b}{c(k^{2}-2)}},
\label{36}
\end{equation}%
where $b<0$ and $c>0$. The elliptic solution given in Eq. (\ref{34}) and Eq.
(\ref{2}) leads to periodic wave function for the Ginzburg-Landau Eq. (\ref{1}) as 
\begin{equation}
\psi (z,t)=\pm \sqrt{\frac{2b}{c(k^{2}-2)}}\,\mathrm{dn}(w(\xi -\xi
_{0}),k)\exp \left[ i(\kappa z-\omega_{s} t+\theta )\right] ,  \label{37}
\end{equation}%
where the inverse width $w$ is given in Eq. (\ref{36}). Let us choose $k=1$
then Eq. (\ref{37}) yields the soliton solution: 
\begin{equation}
\psi (z,t)=\pm \sqrt{-\frac{2b}{c}}\,\mathrm{sech}(w(\xi -\xi _{0}))\exp %
\left[ i(\kappa z-\omega_{s} t +\theta )\right] ,  \label{38}
\end{equation}%
with $w=\sqrt{-b}$, and $b<0$, $c>0$. The latter solution is identical to
the soliton waveform (\ref{32}), found by setting the modulus $k=1$ in the 
\textrm{cn}-type periodic wave (\ref{31}).

Figure 5(a) displays the evolution of the intensity profile of the periodic
wave solution (\ref{26}) of Eq. (\ref{1}) for the values: $\alpha =0.2$,\ $%
\beta =0.1$, $\gamma =-0.4$, $\sigma =-0.2$, $\delta =0.182$, and $u=0.01.$
The intensity profiles of the periodic solutions (\ref{31}) and (\ref{37})
are depited in Figs. 5(b) and 5(c) for the values $\alpha =0.2$,\ $\beta
=0.1 $, $\gamma =0.4$, $\sigma =0.2$, $\delta =-0.182$, and $u=0.01$. We
also choose the position $\xi _{0}$ of the periodic waves at $z=0$ to be
equal to zero. Furthermore, we set the elliptic modulus $k$ to $k=0.9$.\
From these figures, we notice that the three periodic wave solutions exhibit
an oscillating character but they do not have the same shape.

\begin{description}
\item[\textbf{4. Periodic sn(x,k)/(1+cn(x,k)) and tanh(x)/(1+sech(x))
solutions}] 
\end{description}

We have found that Eq. (\ref{22}) has the exact periodic unbounded solution: 
\begin{equation}
F(\xi )=B\,\frac{\mathrm{sn}(w(\xi -\xi _{0}),k)}{1+\mathrm{cn}(w(\xi
-\xi_{0}),k)},  \label{39}
\end{equation}%
Equation (\ref{22}) with Eq. (\ref{39}) allow us to obtain the expression: 
\begin{equation}
\frac{B^{2}w^{2}(1-k^{2})}{\left( 1+\mathrm{cn}(x)\right) ^{2}}+\frac{
B^{2}w^{2}k^{2}\mathrm{cn}^{2}(x)}{\left( 1+\mathrm{cn}(x)\right) ^{2}}+ 
\frac{bB^{2}(1-\mathrm{cn}^{2}(x))}{\left( 1+\mathrm{cn}(x)\right) ^{2}}+ 
\frac{c}{2}B^{4}\frac{\left( 1-\mathrm{cn}^{2}(x)\right) ^{2}}{\left( 1+ 
\mathrm{cn}(x)\right) ^{4}}=Q.  \label{40}
\end{equation}%
Hence, the integration constant is $Q=B^{2}w^{2}/4$, and the inverse width $%
w $ and amplitude $B$ take the form: 
\begin{equation}
w=\sqrt{\frac{2b}{2k^{2}-1}},~~~~B=\pm \sqrt{-\frac{b}{c(2k^{2}-1)}},
\label{41}
\end{equation}%
where $b(2k^{2}-1)>0$ and $c<0$. Using the periodic solution given in Eq. (%
\ref{39}) and Eq. (\ref{2}), we can get a periodic wave function for the
Ginzburg-Landau Eq. (\ref{1}) as 
\begin{equation}
\psi (z,t)=\pm \sqrt{-\frac{b}{c(2k^{2}-1)}}\,\frac{\mathrm{sn}(w(\xi -\xi
_{0}),k)}{1+\mathrm{cn}(w(\xi -\xi _{0}),k)}\exp \left[ i(\kappa
z-\omega_{s} t+\theta )\right] ,  \label{42}
\end{equation}%
where $w$ is given in Eq. (\ref{41}). In the long-wave limit with $k=1$, Eq.
(\ref{42}) leads to the solution: 
\begin{equation}
\psi (z,t)=\pm \sqrt{-\frac{b}{c}}\,\frac{\mathrm{tanh}(w(\xi -\xi _{0}))}{%
1+ \mathrm{sech}(w(\xi -\xi _{0}))}\exp \left[ i(\kappa z-\omega_{s}
t+\theta ) \right] ,  \label{43}
\end{equation}%
with $w=\sqrt{2b}$, and $b>0$, $c<0$. The periodical solution given in Eq. (%
\ref{42}) for parameter $k=0$ reduces to the following periodic solution: 
\begin{equation}
\psi (z,t)=\pm \sqrt{\frac{b}{c}}\,\frac{\mathrm{sin} (w(\xi -\xi_{0}))}{1+%
\mathrm{cos}(w(\xi -\xi _{0}))}\exp \left[ i(\kappa z-\omega_{s} t+\theta )%
\right] ,  \label{44}
\end{equation}%
where $w=\sqrt{-2b}$, and $b<0$, $c<0$.

\begin{description}
\item[\textbf{5. Periodic sn(x,k)/(1+dn(x,k)) and tanh(x)/(1+sech(x))
solutions}] 
\end{description}

Equation (\ref{22}) possesses a periodic solution of the form: 
\begin{equation}
F(\xi )=B\,\frac{\mathrm{sn}(w(\xi -\xi _{0}),k)}{1+\mathrm{dn}(w(\xi -\xi
_{0}),k)}.  \label{45}
\end{equation}%
Then Eq. (\ref{22}) with Eq. (\ref{45}) allow us to obtain the equation: 
\begin{equation}
\frac{B^{2}w^{2}(k^{2}-1)}{k^{2}\left( 1+\mathrm{dn}(x)\right) ^{2}}+\frac{%
B^{2}w^{2}\mathrm{dn}^{2}(x)}{k^{2}\left( 1+\mathrm{dn}(x)\right) ^{2}}+%
\frac{bB^{2}(1-\mathrm{dn}^{2}(x))}{k^{2}\left( 1+\mathrm{dn}(x)\right) ^{2}}%
+\frac{c}{2}B^{4}\frac{\left( 1-\mathrm{dn}^{2}(x)\right) ^{2}}{k^{4}\left(
1+\mathrm{dn}(x)\right) ^{4}}=Q.  \label{46}
\end{equation}%
Hence, the integration constant is $Q=B^{2}w^{2}/4$, and the inverse width $%
w $ and amplitude $B$ take the form: 
\begin{equation}
w=\sqrt{\frac{2b}{2-k^{2}}},~~~~B=\pm k^{2}\sqrt{-\frac{b}{c(2-k^{2})}},
\label{47}
\end{equation}%
where $b>0$ and $c<0$. Using the periodic solution given in Eq. (\ref{45})
and Eq. (\ref{2}), we can get a periodic wave function for the
Ginzburg-Landau Eq. (\ref{1}) as 
\begin{equation}
\psi (z,t)=\pm k^{2}\sqrt{-\frac{b}{c(2-k^{2})}}\,\frac{\mathrm{sn}(w(\xi
-\xi _{0}),k)}{1+\mathrm{dn}(w(\xi -\xi _{0}),k)}\exp \left[ i(\kappa
z-\omega_{s} t+\theta )\right] ,  \label{48}
\end{equation}%
where $w$ is given in Eq. (\ref{47}). In the long-wave limit which
corresponds to $k=1$, the solution in Eq. (\ref{48}) reduces to the wave
function: 
\begin{equation}
\psi (z,t)=\pm \sqrt{-\frac{b}{c}}\,\frac{\mathrm{tanh}(w(\xi -\xi _{0}))}{1+%
\mathrm{sech}(w(\xi -\xi _{0}))}\exp \left[ i(\kappa z-\omega_{s} t+\theta )%
\right] ,  \label{49}
\end{equation}%
with $w=\sqrt{2b}$, and $b>0$, $c<0$. The latter solution coincides with the
wave function (\ref{43}), obtained by setting the modulus $k=1$ in the
periodical solution (\ref{42}).

\begin{description}
\item[\textbf{6. Periodic cn(x,k)/(1+sn(x,k)) solution}] 
\end{description}

We find that Eq. (\ref{22}) has a periodic solution of the form: 
\begin{equation}
F(\xi )=B\,\frac{\mathrm{cn}(w(\xi -\xi _{0}),k)}{1+\mathrm{sn}(w(\xi -\xi
_{0}),k)},  \label{50}
\end{equation}%
Consequently, Eq. (\ref{22}) and Eq. (\ref{50}) lead to equation: 
\begin{equation}
\frac{B^{2}w^{2}}{\left( 1+\mathrm{sn}(x)\right) ^{2}}-\frac{B^{2}w^{2}k^{2}%
\mathrm{sn}^{2}(x)}{\left( 1+\mathrm{sn}(x)\right) ^{2}}+\frac{bB^{2}(1-%
\mathrm{sn}^{2}(x))}{\left( 1+\mathrm{sn}(x)\right) ^{2}}+\frac{c}{2}B^{4}%
\frac{\left( 1-\mathrm{sn}^{2}(x)\right) ^{2}}{\left( 1+\mathrm{sn}%
(x)\right) ^{4}}=Q.  \label{51}
\end{equation}%
As a result, the integration constant is $Q=B^{2}w^{2}(1-k^{2})/4$, and the
inverse width $w$ and amplitude $B$ read: 
\begin{equation}
w=\sqrt{-\frac{2b}{1+k^{2}}},~~~~B=\pm \sqrt{\frac{b(1-k^{2})}{c(1+k^{2})}},
\label{52}
\end{equation}%
where $b<0$, $c<0$ and $0<k<1$. The substitution of periodic solution given
in Eq. (\ref{50}) into Eq. (\ref{2}) yields a periodic wave solution for the
Ginzburg-Landau Eq. (\ref{1}) as 
\begin{equation}
\psi (z,t)=\pm \sqrt{\frac{b(1-k^{2})}{c(1+k^{2})}}\,\frac{\mathrm{cn}(w(\xi
-\xi _{0}),k)}{1+\mathrm{sn}(w(\xi -\xi _{0}),k)}\exp \left[ i(\kappa
z-\omega_{s} t+\theta )\right] ,  \label{53}
\end{equation}%
where $w$ is given in Eq. (\ref{52}). The periodical solution in Eq. (\ref{53}) for parameter $k=0$ reduces to the following periodic solution: 
\begin{equation}
\psi (z,t)=\pm \sqrt{\frac{b}{c}}\,\frac{\mathrm{cos}(w(\xi -\xi _{0}))}{1+%
\mathrm{sin}(w(\xi -\xi _{0}))}\exp \left[ i(\kappa z-\omega_{s} t+\theta )%
\right] ,  \label{54}
\end{equation}%
where $w=\sqrt{-2b}$, and $b<0$, $c<0$.

\begin{description}
\item[\textbf{7. Periodic cn(x,k)/(}$\mathbf{k^{\prime} }$\textbf{+dn(x,k))
solution}] 
\end{description}

We have also found another periodic wave solution for Eq. (\ref{22}) of the
form:%
\begin{equation}
F(\xi )=B\,\frac{\mathrm{cn}(w(\xi -\xi _{0}),k)}{k^{\prime} +\mathrm{dn}%
(w(\xi -\xi _{0}),k)},  \label{55}
\end{equation}
with $k^{\prime}=\sqrt{1-k^{2}}$. Therefore Eq. (\ref{22}) and Eq. (\ref{55}%
) lead to equation: 
\begin{equation}
\frac{B^{2}w^{2}(1-\mathrm{dn}^{2}(x))\left[ 1-k^{2}+k^{\prime} \mathrm{dn}%
(x)\right] ^{2}}{k^{2}\left( k^{\prime} +\mathrm{dn}(x)\right) ^{4}}+\frac{%
bB^{2}\left(k^{2}-1+\mathrm{dn}^{2}(x)\right)} {k^{2}\left(k^{\prime} +%
\mathrm{dn}(x)\right) ^{2}} +\frac{cB^{4}\left( k^{2}-1+\mathrm{dn}%
^{2}(x)\right) ^{2}}{2k^{4}\left(k^{\prime} +\mathrm{dn}(x)\right) ^{4}}=Q.
\label{56}
\end{equation}%
Hence, the integration constant is $Q=B^{2}w^{2}/4$, while the inverse width 
$w$ and amplitude $B$ take the form: 
\begin{equation}
w=\sqrt{\frac{2b}{2-k^{2}}},~~~~B=\pm k^{2}\sqrt{-\frac{b}{c(2-k^{2})}}.
\label{57}
\end{equation}%
Using the periodic solution (\ref{55}) and Eq. (\ref{2}), we can get a
periodic wave function for the Ginzburg-Landau Eq. (\ref{1}) as 
\begin{equation}
\psi (z,t)=\pm k^{2}\sqrt{-\frac{b}{c(2-k^{2})}}\,\frac{\mathrm{cn}(w(\xi
-\xi _{0}),k)}{k^{\prime} +\mathrm{dn}(w(\xi -\xi _{0}),k)}\exp \left[
i(\kappa z-\omega_{s} t+\theta )\right] ,  \label{58}
\end{equation}
where $b>0$ and $c<0$.

In Fig. 6(a), we present the evolution of the intensity profile of the
periodic wave solution (\ref{42}) of the model (\ref{1}) for the values: $%
\alpha =0.2$,\ $\beta =0.1$, $\gamma =-10$, $\sigma =-5$, $\delta =0.014$,
and $u=0.01.$ We also show the intensity profile of the periodic solution (%
\ref{48}) in Fig. 6(b) for the values $\alpha =0.2$,\ $\beta =0.1$, $%
\gamma=-0.326$, $\sigma =-0.163$, $\delta =0.083$, and $u=0.01$ as well as
the periodic wave (\ref{53}) for the values $\alpha =0.2$,\ $\beta =0.1$, $%
\gamma =6.4$, $\sigma =3.2$, $\delta =-0.068$, and $u=0.01$ in Fig. 6(c).
The results for the periodic wave solution (\ref{58}) is displayed in Fig.
6(c), where we used the same model parameters as in Fig. 6(b). Here the
position $\xi _{0}$ of the periodic waves at $z=0$ is taken as $\xi _{0}=0$.
In addition, the value of elliptic modulus $k$ is selected as $k=0.6$.\ We
clearly see from these figures that the four periodic wave solutions show
oscillating behavior with distinct shapes.

Figure 7(a) illustrates the evolution of the intensity profile of the dark
soliton solution (\ref{27}) for the parameter values: $\alpha =0.2$,\ $\beta
=0.1$, $\gamma =-0.4$, $\sigma =-0.2$, $\delta =0.182$, and $u=0.01$. The
intensity profile of the bright pulse solution (\ref{32}) is shown in Fig.
7(b) for the values: $\alpha =0.2$,\ $\beta =0.1$, $\gamma =0.4$, $\sigma
=0.2$, $\delta =-0.182$, and $u=0.01.$ Additionally, we present in Fig. 7(c)
the intensity profile of the dark waveform solution (\ref{43}) for the same
parameter values as those in Fig. 7(a). These results indicate that
dark-type soliton structures may exist in two distinct functional forms
within the framework of the CGLE (\ref{1}). To the best of our knowledge,
the localized and periodic waves presented above for the Ginzburg-Landau Eq.
(\ref{1}) are reported here for the first time.

\section{Stability analysis for solutions of Ginzburg-Landau equation}

We consider in this section the analytical stability analysis of solutions
for the Ginzburg-Landau equation based on the theory of nonlinear dispersive
waves in nonlinear optics \cite{HK,KH}. For our purpose, we develop the
dynamics of dispersive waves in the form: 
\begin{equation}
\psi (z,t)=U(\omega)\exp[ i\Theta(z,t)],  \label{59}
\end{equation}
where the amplitude $U(\omega)$ and phase $\Theta(z,t)$ are real functions.
The frequency $\omega=\omega(z^{\prime},t^{\prime})$ is a slow varying
function depending on slow variables $z^{\prime}=\varepsilon z$ and $%
t^{\prime}=\varepsilon t$ where $\varepsilon\ll 1$ is a small dimensionless
parameter. We can conciser the frequency $\omega$ of nonlinear dispersive
waves as slow varying function of $z^{\prime}$ and $t^{\prime}$ because the
frequency $\omega_{s}$ in Eq. (\ref{2}) is constant. The wave number $%
k(\omega)$ and frequency $\omega$ of the nonlinear dispersive waves are
given by the equations, 
\begin{equation}
k(\omega)=\frac{\partial \Theta}{\partial z}, ~~~~\omega=-\frac{\partial
\Theta}{\partial t}.  \label{60}
\end{equation}
Equation (\ref{60}) yields the relations $\Theta_{zt}=k_{t}$ and $%
\Theta_{tz}=-\omega_{z}$, which lead to equation, 
\begin{equation}
\frac{\partial \omega}{\partial z} +\frac{\partial k(\omega)}{\partial t}=0.
\label{61}
\end{equation}
Equations (\ref{59}) and (\ref{1}) lead to the following coupled equations:%
\begin{equation}
U_{tt}-\frac{2\beta\omega}{\alpha} U_{t} -\left(\frac{\kappa}{\alpha}%
+\omega^{2} +\frac{\beta\omega_{t}}{\alpha}\right)U +\frac{\gamma}{\alpha}
U^{3}=0,  \label{62}
\end{equation}
\begin{equation}
U_{tt}-\frac{1}{\beta}U_{z}+\frac{2\alpha \omega}{\beta}U_{t} +\left(\frac{%
\delta}{\beta}-\omega^{2} +\frac{\alpha\omega_{t}}{\beta}\right)U +\frac{%
\sigma}{\beta}U^{3}=0.  \label{63}
\end{equation}%
Equations (\ref{62}) and (\ref{63}) are self consistent when the following
equations are satisfied: 
\begin{equation}
-\frac{2\beta\omega}{\alpha} U_{t}= -\frac{1}{\beta}U_{z}+\frac{2\alpha
\omega}{\beta}U_{t},  \label{64}
\end{equation}
\begin{equation}
k(\omega)=-\frac{\alpha\delta}{\beta} -\frac{1}{\beta}(\alpha^{2}+\beta^{2})%
\omega_{t},  \label{65}
\end{equation}
with the constraint as $\beta\gamma=\alpha\sigma$. We can also write Eq. (%
\ref{64}) as 
\begin{equation}
U_{z}=\frac{2}{\alpha}(\alpha^{2}+\beta^{2})\omega U_{t},  \label{66}
\end{equation}
which yields the equation for frequency: 
\begin{equation}
\omega_{z}=\frac{2}{\alpha}(\alpha^{2} +\beta^{2})\omega\omega_{t}.
\label{67}
\end{equation}
Equations (\ref{61}) and (\ref{67}) lead the following equation for the wave
number $k(\omega)$: 
\begin{equation}
k_{t}=-\frac{2}{\alpha}(\alpha^{2}+\beta^{2}) \omega\omega_{t}.  \label{68}
\end{equation}
We also have equation for $k_{t}$ by differentiation of Eq. (\ref{65}): 
\begin{equation}
k_{t}=-\frac{1}{\beta}(\alpha^{2}+\beta^{2})\omega_{tt}.  \label{69}
\end{equation}
Equations (\ref{68}) and (\ref{69}) lead to equation for the frequency $%
\omega$ as $(\omega^{2})_{t}=(\alpha/\beta)\omega_{tt}$. Integration of this
equation leads to the following differential equation for the frequency $%
\omega$: 
\begin{equation}
\omega_{t}=\frac{\beta}{\alpha}\omega^{2} +\frac{\beta}{\alpha}%
\omega_{0}^{2},  \label{70}
\end{equation}
where $\omega_{0}=\omega_{0}(z)$ is an arbitrary function of $z$. Thus, Eqs.
(\ref{65}) and (\ref{70}) lead to the wave number $k(\omega)$ as 
\begin{equation}
k(\omega)=-\frac{\alpha\delta}{\beta}-\frac{1}{\alpha}(\alpha^{2}+%
\beta^{2})(\omega^{2}+\omega_{0}^{2}).  \label{71}
\end{equation}
Integration of Eq. (\ref{70}) with condition $\omega=0$ for $t=t_{0}$ yields
the solution: 
\begin{equation}
\omega=\omega_{0}(z)\tan\left(\frac{\beta}{\alpha} \omega_{0}(z)(t-t_{0})%
\right).  \label{72}
\end{equation}
We note that the frequency $\omega=\omega(z^{\prime},t^{\prime})$ is the
function of slow variables $z^{\prime}=\varepsilon z$ and $%
t^{\prime}=\varepsilon t$. This form for the slow varying function $\omega$
follows from Eq. (\ref{72}): 
\begin{equation}
\omega(z^{\prime},t^{\prime})=\varepsilon\Omega (z^{\prime})\tan\left(\frac{%
\beta}{\alpha} \Omega(z^{\prime})(t^{\prime}-t_{0}^{\prime})\right),
\label{73}
\end{equation}
where we have presented an arbitrary function $\omega_{0}(z)$ as $%
\omega_{0}(z)=\varepsilon\Omega(z^{\prime})$, and initial time is $%
t_{0}^{\prime}=\varepsilon t_{0}$.

In the first order to small parameter $\varepsilon$ Eqs. (\ref{62}) and (\ref%
{63}) lead to nonlinear differential equation: 
\begin{equation}
U_{z}-2\alpha\omega U_{t}=\delta U+\sigma U^{3}.  \label{74}
\end{equation}
The characteristic system of equations for the differential equation (\ref{74}) is given as 
\begin{equation}
\frac{dz}{d\tau}=1,~~~~\frac{dt}{d\tau}=-2\alpha\omega,  \label{75}
\end{equation}
\begin{equation}
\frac{dU}{d\tau}=\delta U+\sigma U^{3}.  \label{76}
\end{equation}
The trajectory of characteristic Eqs. (\ref{75}) and (\ref{76}) is defined
by the following equations: 
\begin{equation}
t=t(z),~~~~U=U(z,t(z)).  \label{77}
\end{equation}
The solution of Eq. (\ref{76}) with initial condition $(U)_{z=z_{0}}=U_{0}$
and relation $dz=d\tau$ is given as 
\begin{equation}
U(z)=U_{0}\sqrt{\frac{\delta} {(\delta+\sigma U_{0}^{2})\exp[%
-2\delta(z-z_{0})] -\sigma U_{0}^{2}}}.  \label{78}
\end{equation}
We can use the solution for amplitude of dispersive waves given in Eq. (\ref{78}) for stability analysis of solutions of the Ginzburg-Landau equation.

We suppose that the solutions of nonlinear wave equations are stable when
the optical pulses not radiate the dispersive waves. Such situations exist
when the characteristic system of equations has not solutions \cite{HK,KH}.
The stability analysis for solutions of Ginzburg-Landau equation is based on
Eq. (\ref{78}) which demonstrates that intensity of the dispersive waves
depends on two parameters $\delta$ and $\sigma$.

We conciser below two cases for parameters $\delta$ and $\sigma$: 1. $\delta<0$ and $\sigma>0$, 2. $\delta<0$ and $\sigma\leq 0$. Moreover, in the
first case we also assume that $\delta+\sigma U_{0}^{2}<0$ because the
intensity $U_{0}^{2}$ of initial dispersive waves is relatively small. Let
us allow the propagation distances $z-z_{0}$ for dispersive waves (in the
above two cases) satisfying the condition: $|\delta+\sigma U_{0}^{2}| \exp[%
-2\delta(z-z_{0})]\gg |\sigma|U_{0}^{2}$ where $\delta<0$. Then Eq. (\ref{78}%
) with this condition leads to the following equation for intensity of
dispersive waves: 
\begin{equation}
U(z)^{2}= \frac{|\delta| U_{0}^{2}} {|\delta+\sigma U_{0}^{2}|}\exp[%
-2|\delta|(z-z_{0})].  \label{79}
\end{equation}
Hence, we have found that in the first and second cases for parameters $%
\delta$ and $\sigma$ the intensity of dispersive waves is decreasing
exponentially. We define the solutions of CGLE (\ref{1}) as quasi-stable
when the dispersive waves are decreasing exponentially with propagation
distances.

In the context of passive mode locking lasers (see Appendix A) the first
regime with $\delta<0$ and $\sigma>0$ occurs when the following inequalities
are satisfied: 
\begin{equation}
\alpha_{c}+\alpha_{0}>g_{c},,~~~~\alpha_{0}P_{s}^{-1} > \alpha_{2}.
\label{80}
\end{equation}
The second regime with $\delta<0$ and $\sigma\leq 0$ yields the conditions
for parameters of lasers as 
\begin{equation}
\alpha_{c}+\alpha_{0}>g_{c},~~~~\alpha_{0}P_{s}^{-1}\leq \alpha_{2}.
\label{81}
\end{equation}
Here the parameters $\alpha_{c}$, $\alpha_{0}$ and the saturation power of
the absorber $P_{s}$ are connected with the cavity-loss parameter of passive
mode locking fiber lasers. In these inequalities the parameter $\alpha_{2}$
describes two-photon absorption and $g_{c}$ is the saturated gain averaged
over the cavity.

Let us consider the stability of periodic elliptic solution \textrm{dn(x,k)}
and soliton solution \textrm{sech(x)} given by Eqs. (\ref{37}) and (\ref{38}%
) with parameters $b<0$ and $c>0$. In the case when inverse velocity $u$ is
zero we have by Eq. (\ref{8}) that $b=\delta /\beta $ and $c=\sigma /\beta $%
. It follows from Eqs. (\ref{a3}) and (\ref{a4}) (see Appendix A) that $%
\beta =g_{c}T_{2}^{2}/2>0$ which leads to conditions $\delta <0$ and $\sigma
>0$ for existing of the periodic elliptic solution \textrm{dn(x,k)} and
soliton solution \textrm{sech(x)}. Thus, the periodic elliptic solution 
\textrm{dn(x,k)} and soliton solution \textrm{sech(x)} are quasi-stable. The
same result takes place for relatively small inverse velocity $u$ satisfying
to Eq. (\ref{20}).

\section{Conclusion}

We have examined the existence of localized and periodic waves in
dissipative systems wherein the pulse propagation is modeled by the cubic
complex Ginzburg-Landau equation. Using the ansatz method, we showed that
two new types of kink and antikink solitons with distinct functional forms
can exist in the nonlinear medium with fixed inverse velocity. A very
physically relevant property is that the two kink-type waves have equal
inverse velocity which is found to depend on parameters of fiber laser
system, such as the group delay dispersion and the spectral filtering
coefficients. We also demonstrated that the nonlinear medium supports
various novel periodic waves in different forms, such as \textrm{sn}, 
\textrm{cn}, \textrm{dn}, and their rational forms, thus illustrating the
richness of the fiber laser medium. We have also found that these periodic
waves degenerate into different bright and dark soliton pulses in the
long-wave limit. We also have presented the analytical stability analysis of
solutions for the Ginzburg-Landau equation based on the theory of nonlinear
dispersive waves in nonlinear optics. Finally, we hope that the precise
analytical forms of the obtained nonlinear waves may be profitably exploited
to the design of fiber laser systems.

\appendix

\section{Passive mode locking based on Ginzburg-Landau equation}

The theory of passive mode locking fiber lasers is based on the
Ginzburg-Landau equation. In this model the cavity-loss parameter $\tilde{%
\alpha}$ include the intensity dependence of loss produced by the saturable
absorber: 
\begin{equation}
\tilde{\alpha}=\alpha _{c}+\alpha _{0}(1+P_{s}^{-1}\left\vert \psi
\right\vert ^{2})^{-1}\approx \alpha _{c}+\alpha _{0}-\alpha
_{0}P_{s}^{-1}\left\vert \psi \right\vert ^{2},  \label{a1}
\end{equation}%
where $P_{s}$ is the saturation power of the absorber, assumed to be much
larger than the peak power levels associated with optical pulses circulating
inside the laser cavity. Thus, we assume here the condition as $%
P_{s}^{-1}\left\vert \psi \right\vert ^{2}\ll 1$. The Ginzburg-Landau
equation in this fiber laser model \cite{Agraw} is 
\begin{equation}
i \frac{\partial \psi }{\partial z}-\frac{1}{2} (\beta_{2}+i\beta_{1}) \frac{%
\partial ^{2}\psi}{\partial t^{2}} +(\gamma-i\sigma) \left\vert \psi
\right\vert^{2}\psi-\frac{i}{2}g_{0}\psi=0,  \label{a2}
\end{equation}
where the parameters $\beta_{1}$, $\sigma$ and $g_{0}$ are 
\begin{equation}
\beta_{1}=g_{c}T_{2}^{2},~~~~ \sigma=\frac{1}{2}(\alpha_{0}P_{s}^{-1}-%
\alpha_{2}),~~~~ g_{0}=g_{c}-\alpha_{c}-\alpha_{0}.  \label{a3}
\end{equation}
Here the parameter $T_{2}$ is related to the gain bandwidth as $%
T_{2}=\Omega_{g}^{-1}$, the parameter $\alpha_{2}$ is connected with
two-photon absorption and $g_{c}$ is the saturated gain averaged over the
cavity. The Ginzburg-Landau equation (\ref{1}) has the same form as Eq. (\ref%
{a2}) with parameters $\alpha$, $\beta$ and $\delta$ defined as 
\begin{equation}
\alpha=-\frac{1}{2}\beta_{2},~~~~\beta=\frac{1}{2}\beta_{1},~~~~ \delta=%
\frac{1}{2}g_{0}.  \label{a4}
\end{equation}

\end{document}